Hard-sphere model of the B2 → B19' phase transformation, and its application to predict the B19 structure in NiTi alloys and the B19 structures in other binary alloys

Cyril Cayron*

* Laboratory of Thermo Mechanical Metallurgy (LMTM), PX Group Chair, Rue de la Maladière 71b, 2000 Neuchâtel, Switzerland, cyril.cayron@epfl.ch

**Abstract:** The pseudoelastic and pseudoplastic properties of NiTi alloys result from the closeness of the structures between the B2 cubic austenite and the B19' monoclinic martensite, and the facility to transform one into each other. Until now, the paths followed by the atoms during the B2 → B19' transformation were imagined as independent shears and shuffles. Here, we propose a simplified hard-sphere atomistic model of phase transformation decomposed into three distinct types of atomic movements: a contraction along the $[100]_{B2}$ axis resulting from the formation of Ti-Ti bonds, an angular distortion between the $[001]_{B2}$ and $[1\bar{1}0]_{B2}$ directions, and an rotation of the Ti atoms around the Ni atoms that also establishes new Ti-Ti bonds. The model's inputs are the Ti and Ni atomic or ionic diameters and the monoclinic angle $\beta$. The outputs are the lattice parameters of the B19' phase and the atomic positions. The results are remarkably close to those reported in the literature from X-ray diffraction experiments or DTF simulations. The hard-sphere model permits a better understanding of the B19' structure and the way it is inherited from its parent B2. It explains the change of enthalpy by the formation of short Ti-Ti bonds in B19'. It is also shown that the value of the monoclinic angle $\beta$ close to 97.9° corresponds to the highest molar volume among all the possible hard-sphere monoclinic B19' structures; which suggests that it could be a consequence of a maximization of the vibrational entropy. The hard-sphere model was applied in the special case of absence of monoclinicity to predict the B19 structure. The calculations do not agree well with the B19 structure reported in NiTi alloys, probably because of the phase only forms when the alloys contain a high content of copper. They are however in excellent agreement with the B19 structures reported in other binary alloys, such as in AuTi, PdTi, and AuCd. It is concluded that the B19' phase in NiTi alloys and the B19 phase in AuTi, PdTi, and AuCd alloys can be considered as "simple" hard-sphere structures.

# 1 Introduction

NiTi are the most studied and used shape memory alloys since their discovery more than sixty years ago [1]. The knowledge of the B19' structure is important to a get a better understanding of the B2 → B19' martensitic transformation and B19' → B19' variant reorientation at the origin of the pseudoelasticity and pseudoplasticity, respectively. The B19' phase was studied by X-ray and neutron diffraction by many groups in the past [2]-[5], but some structures and space groups initially proposed were rough or incorrect. The long and incremental history of this research is detailed in Ref.[6]. Michal and Sinclair in 1981 [3] were the first to discover the correct monoclinic space group $P2_1/m$ (n°11). The lattice parameters[1] they proposed (a = 2.885 Å, b = 4.120 Å, c = 4.622 Å, and $\beta$ = 96.8°) are quite

---

[1] Please note that the B19' structure can be reported with two equivalent crystallographic conventions depending on the choice of the monoclinic angle $\beta$ or $\gamma$. The interchange between them is simply as follows: lattice (a, b, c, $\beta$), direction [x, y, z], plane (h, k, l) ↔ lattice (a, c, b, $\gamma$), direction [x, z, y], plane

close to those determined ten years earlier by Otsuka *et al.* [4]. In 1985, Kudoh *et al.* [5] refined further the $P2_1/m$ structure and obtained *a* = 2.898 Å, *b* = 4.108 Å, *c* = 4.646 Å, and $\beta$ = 97.78°. The difference between the different studies may also come from to the difference of compositions of the alloys. In some studies the alloy was made of 50-50 atomic ratio of Ni and Ti [3], while in others the ratio of nickel was lower, for example in Ref. [5]. The effect of composition and temperature on the lattice parameters of B19' has been investigated in details by X-ray diffraction by Prokoshkin *et al.* [7]. For an exact 50-50 ratio, at room temperature, the lattice parameters they measured are *a* = 2.904 Å, *b* = 4.121 Å, *c* = 4.649 Å, and $\beta$ = 97.9°, which are very close to those determined by Kudoh *et al.* [5]. Another reason for the difference of the lattice parameters reported in the literature is the broadening of some diffraction peaks in the diffractograms. They may be the consequence of the elastic field required for the accommodation between the parent B2 and the martensite, or between the different martensitic variants [8].

Some researchers used first principle calculations of electronic density of state (DOS) to determine and compare the ground state energies of the different B19' structures reported in the literature. Sanati *et al.* [9] calculated that the structure given by Kudoh *et al.* [5] has the lowest total energy at zero temperature. Fukuda *et al.* [10] showed that the Fermi energy decreases along a sequence of martensitic transformations B2 → R →B19'. Note that the phase R is a trigonal phase (initially thought rhombohedral) and is formed in some NiTi alloys that contain additional elements such as Fe and Al [6]. For the last two decades, the DOS calculations have been performed by using the Density Functional Theory (DFT). Hang *et al.* [11] found that the most stable B19' structure should have a monoclinic angle $\beta = 107°$ instead of 98°. With this special angle, new symmetries appear and the monoclinic structure become a base-centred orthorhombic (BCO) superstructure, later identified as B33 [12]. The high monoclinic angle obtained by Hang *et al.* [11] has been confirmed by other DFT calculations by Hatcher *et al.* [13] and Vishnu and Strachan [14] who found $\beta_{DFT} \approx 109°$ and $\beta_{FT} \approx 106°$, respectively. Since the B33 structure has never been observed, Huang *et al.* [11] proposed that the actual B19' structure ($\beta_{B19'} \approx 98°$) exists in place of B33 ($\beta_{DFT} \approx 107°$) because of the residual stresses. This argument is however controversial because it does not hold for samples recrystallized at high temperatures. Another explanation of the better stability of B19' compared to B33 was proposed by Ishii [15]. He demonstrated that the energy required for the elastic accommodation between the parent B2 matrix and the martensite is lower for B19' than for B33 (mainly because of the lower monoclinic angle). He thus concluded that B33 forms but reaches instability and changes into B19' in order to release the internal stresses. Imagining B19' as a B33 phase distorted by back stresses allowed him to predict an "equilibrium" monoclinic angle of 99.3°, not far from the actual angle $\beta_{B19'} \approx 98°$. To our opinion, however, the hypothetical existence of a B33 phase predicted by DFT should not be overestimated. The set of lattice parameters found for B19' by DFT deviates by 4% from the exact relation $\cos\beta = -b/2a$ [11] required to form the BCO (B33) symmetries. It is also probable that the entropy plays a major contribution to the Gibbs energy at room temperature and makes the B19' structure deviates from its expected ground state. We do not think that B33 should be considered as a key phase to understand the B19' structure.

The mechanism and atomic trajectories during the B2 to B19' transformation are unclear. According to Otsuka and Ren [4], the phase change can be schematically explained in a two-step sequence B2→ B19 →B19'. The first step is obtained by a lattice stretching along the $[001]_{B2}$, $[110]_{B2}$ and $[1\bar{1}0]_{B2}$ axes, and is coupled with a shuffle along the $[1\bar{1}0]_{B2}$ direction for the atoms located in the $(110)_{B2}$

---

(*h*, *l*, *h*). In the present paper, we use the *β*-convention recommended in the International Table for Crystallography.

plane. The second step is the monoclinic distortion that transforms the 90° angle of B19 into the monoclinic angle $\beta_{B19'} \approx 98°$. This step is described as a lattice shear on the $(001)_{B2}$ plane along the $[1\bar{1}0]_{B2}$ direction. Since the atomic positions after this shear do not however correspond to those deduced from X-ray diffraction, additional shuffles are required. In this description, the shears and shuffles are independent, but it should not be so because the atomic displacements and the lattice distortion are highly interconnected in most of the displacive phase transformations [16]. Kibey *et al.* [17] used DFT to determine the atomic trajectories during the B2→ B19 transformation by imposing the final lattice parameters of B19 and by assuming that the transient orthorhombic states are intermediate between B2 and B19 with lattice parameters that depend linearly on $a_{B2}$ and $a_{B19}$, $b_{B19}, c_{B19}$ with a unique degree of evolution λ that varies from 0 to 1. The calculated energy landscape represented in the (λ, η) subspace, where η is the normalized shuffle amplitude, does not help however to get any idea of the atomic trajectories in the real 3D space. Besides this issue, the study was done only for B19, but not for B19'. More recently, Li *et al.* [18] used the Modified Embedded Atom Method (MEAM) to study the transformation in a quite large numerical single crystal containing one million atoms. The formation of junction planes between the twin-related domains of martensite could be simulated. In addition, the authors proposed a simple atomistic model of B2 → B19' transformation by considering that the Ni and Ti atoms in the $(110)_{B2}$ plane form a distorted 2D hexagonal lattice. They thus suggest that the B2 → B19' transformations result from a kind of body centred cubic (bcc) to hexagonal close packed (hcp) Burgers mechanism. Their proposition relies on the structural proximities of the B19 and hcp structures (B19 is sometimes called "ordered hcp"), but the model remains speculative and does not give quantitative predictions of the lattice distortion and final atomic positions.

The aim of the manuscript is to propose a simple hard-sphere model of B2→ B19' transformation. We have already elaborated such hard-sphere models for transformation in steels between face centred cubic (fcc) austenite and bcc martensite [19], and more generally for the martensitic transformations between the fcc, bcc and hcp phases [20]. It can seem illusory to build a similar hard-sphere model for a phase transformation between intermetallic compounds in which the some directional covalent bonds are expected, but a spherical symmetry of the atoms is not unrealistic if we consider that the number of valence electrons per atom in NiTi, e/a = 0, as reported in Ref.[6]. Actually, we will show that the hard-sphere assumption permits to satisfactorily determine the lattice parameters and atomic positions of the final B19' structure. The only input will be the Ti and Ni atomic diameters and the value of the monoclinic angle. The hard-sphere B19' structure will be shown to be very close to that already determined from X-ray diffraction and from DTF. It will be also shown that the value of the monoclinic angle $\beta \approx 98°$ initially used as an input actually corresponds to the maximum volume change among of the hard-sphere B19' structures that could be obtained by varying $\beta_{B19'}$. The model will be also applied in the specific cases where $\beta = 90°$ to predict the structure of the B19 phase in NiTi and other binary B19 compounds. The predictions of the B19 structures will be shown excellent for B19 phase in AuCd and PtTi alloys, but less good for the B19 phase in NiTi for reasons that will be discussed.

## 2 Hard-sphere model

### 2.1 Inputs

We make the simple assumption that the Ti and Ni atoms are spheres that do not interpenetrates and that come into contact when attractive bonds link them together. This way of considering crystal structures and transformations between them is quite "old-fashion" in comparison with the incredible capabilities of DFT simulations, but, when it is applicable, it has the enormous advantage to capture the most important features of the phase transformation and give a global understanding of the mechanism. In this hard-sphere model, two values are required: the diameters of the Ti and Ni atoms, which are respectively the large and small atoms in the structure. We start by fixing the size of Ti, the largest atom. From the room temperature lattice parameters of metallic Ti in hcp state the diameter should be $d_{Ti} = a_{Ti} = 2.95$ Å. However, since in all the B19' structures proposed in the literature the vector $\boldsymbol{a}_{B19'}$ that points between two Ti atoms has a length slightly lower, close to 2.9 Å, we fixed the diameter of the Ti atoms at $d_{Ti} = a_{B19'} = 2.9$ Å. The diameter of the Ni atoms cannot be determined as directly as for Ti atoms. If it is deduced from the metallic Ni in fcc state with $a_{Ni} = 3.52$ Å, the diameter would be $d_{Ni} = \frac{\sqrt{2}}{2} a_{Ni} = 2.49$ Å. If it is calculated from the Ti-Ni bond in the B2 phase, the diameter would be $d_{Ni} = \sqrt{3} a_{B2} - d_{Ti} = 2.31$ Å, which is significantly lower, and actually corresponds to Ni atoms involved in covalent bonds. Diameters of covalent Ni atoms are indeed reported between $d_{Ni} = 2.20$ Å and $d_{Ni} = 2.30$ Å depending on the source of information [21][22]. After manually fitting the outputs of the model with the X-ray diffractograms reported in the literature (the details will be given in section 3.1), we decided to fix the Ni diameter to $d_{Ni} = 2.27$ Å. The last parameter in the model is the monoclinic angle $\beta$. The value was initially chosen at $\beta_{B19'} = 96.8°$ as reported in Ref. [6], but, as it will be detailed in section 3.1, we realized that $\beta_{B19'} = 97.9°$, a value close to $\beta_{B19'} = 97.8°$ given by Kudoh et al. [5], gives better agreement with the X-ray diffractograms reported in the literature. In summary, the inputs of the hard-sphere model are the three parameters $d_{Ti} = 2.9$ Å, $d_{Ni} = 2.27$ Å, and $\beta_{B19'} = 97.9°$. The atomic diameters of Ti and Ni have a physical meanings (even simplistic), and even if the value of $\beta_{B19'}$ is an input, we will show that it is special and it could probably be determined *a posteriori*.

## 2.2 Hard sphere packing in the first layer of $(110)_{B2}$

The model is built as follows. From the B2 unit cell shown in Fig. 1a, a B2 supercell based on the vectors $[001]_{B2}$, $[110]_{B2}$ and $[\bar{1}10]_{B2}$ is introduced as represented in Fig. 1b. These vectors will become the B19' vectors $\boldsymbol{a}_{B19'}$, $\boldsymbol{b}_{B19'}$ and $\boldsymbol{c}_{B19'}$, as in Otsuka and Ren's model [6]. The associated correspondence matrix is

$$\boldsymbol{C}^{B2 \rightarrow B19'} = \begin{pmatrix} 0 & 1 & -1 \\ 0 & 1 & 1 \\ 1 & 0 & 0 \end{pmatrix} \quad (1)$$

It is the matrix we already used on the correspondence theory to calculate the twins between the B19' variants [8]. The B2 supercell is tetragonal. After stretching it gives the B19 structure, and after angular distortion it gives the B19' structure. Let us consider the atomic displacements that could be compatible with such lattice transformations. We consider the $(110)_{B2}$ plane of the parent phase; it will become the $(010)_{B19'}$ plane of martensite. In the first layer of this plane, the vector $\overrightarrow{OA} = [001]_{B2}$ between the two Ti atoms in positions O and A is shorten from $OA = a_{B2} = 3.01$ Å to $OA' = a_{B19'} = d_{Ti} = 2.9$ Å, and simultaneously tilted such that the angle between $[001]_{B2}$ and $[\bar{1}10]_{B2}$ initially at 90° becomes $\beta_{B19'} = 97.9°$. These two movements are noted 1 and 2 in Fig. 1c. In this model, the Ti atom in position O is fixed. The movement 1 establishes a contact between the Ti atoms in positions O and A. The same movement occurs for the Ti atoms in C and D in order to preserve the lattice structure. During the movements 1 and 2, the Ni atom in position E, initially in contact with the Ti atoms in O, A, C, and D, loses contact with the Ti atom in D, but keeps contact with the other three Ti atoms, as shown in Fig. 1d. Its final position E' is ambiguously determined by geometry. The distance between the Ti atoms in O and C, initially $\sqrt{2} a_{B2}$ is elongated to become $c_{B19'}$. During these displacements, the distance between the Ti and Ni atoms that remain in contact is $d_{TiNi} = \frac{d_{Ni} + d_{Ti}}{2}$.

The calculations in an orthonormal reference basis $\mathcal{B}_⊞ = \left([001]_{B2}, \frac{1}{\sqrt{2}}[110]_{B2}, \frac{1}{\sqrt{2}}[\bar{1}10]_{B2}\right)$ show that

$$c_{B19'} = \sqrt{k} \sin \beta_{B19'} - d_{Ti} \cos \beta_{B19'} \tag{2}$$

with $k = d_{Ni}(d_{Ni} + 2d_{Ti})$. The coordinates of the Ni atom in E' written in $\mathcal{B}_⊞$ are

$$E'_{/\mathcal{B}_⊞} = \left(\frac{-\sqrt{k}\cos\beta_{B19'} + d_{Ti}\sin\beta_{B19'}}{2}, 0, \frac{d_{Ti}\cos\beta_{B19'} + \sqrt{k}\sin\beta_{B19'}}{2}\right) \tag{3}$$

They can be written in the conventional crystallographic basis $\mathcal{B}_{B19'} = (\boldsymbol{a}_{B19'}, \boldsymbol{b}_{B19'}, \boldsymbol{c}_{B19'})$ by using the coordinate transformation matrix, also called structure tensor:

$$[\mathcal{B}_⊞ \to \mathcal{B}_{B19'}] = \begin{pmatrix} a_{B19'}\sin\beta_{B19'} & 0 & 0 \\ 0 & b_{B19'} & 0 \\ a_{B19'}\cos\beta_{B19'} & 0 & c_{B19'} \end{pmatrix} \tag{4}$$

The value of $b_{B19'}$ is not yet known now, but it has no influence on the atoms of the first layer. The coordinates of E' given by $E'_{/\mathcal{B}_{B19'}} = [\mathcal{B}_⊞ \to \mathcal{B}_{B19'}]^{-1} E'_{/\mathcal{B}_⊞}$ are

$$E'_{/\mathcal{B}_{B19'}} = \left(\frac{d_{Ti} - \sqrt{k}/\tan\beta_{B19'}}{2d_{Ti}}, 0, \frac{k/\sin\beta_{B19'}}{2k\sin\beta_{B19'} - 2d_{Ti}\sqrt{k}\cos\beta_{B19'}}\right) \tag{5}$$

The calculation gives $c_{B19'} = 4.638$ Å and $E' = (0.6024, 0, 0.4658)$. Note that these values are close to those given by Kudoh *et al.* [5], i.e. $c_{B19'} = 4.646$ Å and $E' = (0.6196, 0, 0.4588)$.

## 2.3 Hard-sphere packing in the second layer of $(110)_{B2}$

For the atoms of the second layer of the $(110)_{B2}$ plane, two kinds of hard sphere stacking can be imagined: (i) one in which the Ni atom initially in position B comes into contact with the Ti atoms in O and A' and with the Ni atom in E', or (ii) one in which the Ti atoms, initially in F and G in B2, keep contact and maintain their $\beta_{B19'}$ angle established during the movements 1 and 2, and in addition come into contact with the Ni atom in E', as shown in Fig. 1e,f. After investigating these two possibilities, we came to conclude that only the stacking (ii) gives the correct B19' structure. The rotational movement of the Ti atoms is labeled 3 in Fig. 1e. The final position of the Ti atoms that were initially in F and G in B2 are given by the points F' and G' shown in Fig. 1f. The Ni atom initially in position B in the B2 structure keeps contact with these Ti atoms during their displacements from F to F' and G to G', and it takes the place B' in B19'.

The atomic displacements resulting for the three movements are represented in 3D in Fig. 2, with a top view along the direction [001]<sub>B2</sub> in Fig. 2a,b, and a side view in Fig. 2c,d. At completion, they permit to directly form B19' without any additional shear or shuffle.

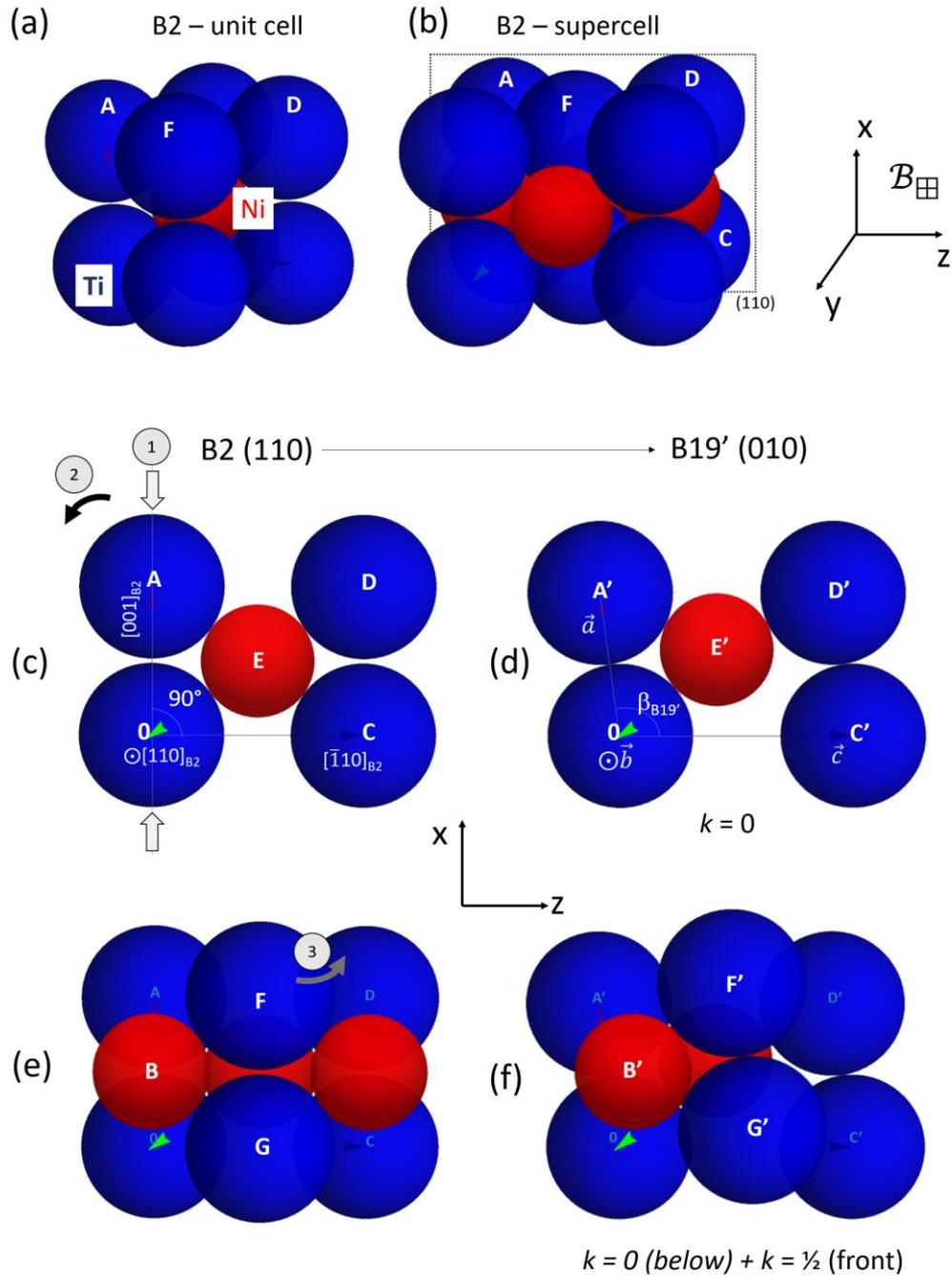

Fig. 1. Atomistic model of the B2→ B19' phase transformation. (a) B2 unit cell formed by one Ti atom at the corners (in blue) and one smaller Ni atom in the centre (in red). (b) B2 supercell formed by the vectors [110], [1̄10], [001]. This supercell contains two Ti atoms and two Ni atoms. (c) The first layer of the plane (110)$_{B2}$ is transformed into (d) the first layer of the plane (010)$_{B19'}$ ($k = 0$). Note that the Ti atoms in O and A come into contact with each other by the movement 1, and that the direction OA rotates from $\beta_i$ = 90° (B2) to $\beta_f$ = 97.9° (B19') by the movement 2. The Ni atom in E keeps contact with the Ti atoms in O and A. (e) The second layer of the plane (110)$_{B2}$ is transformed into (f) the second layer of the plane (010)$_{B19'}$ ($k = ½$). The Ti atom in F comes into contact with the Ti atom of the first layer in D' by the movement 3. The green arrow represents the b$_{B19'}$ axis perpendicular to the page.

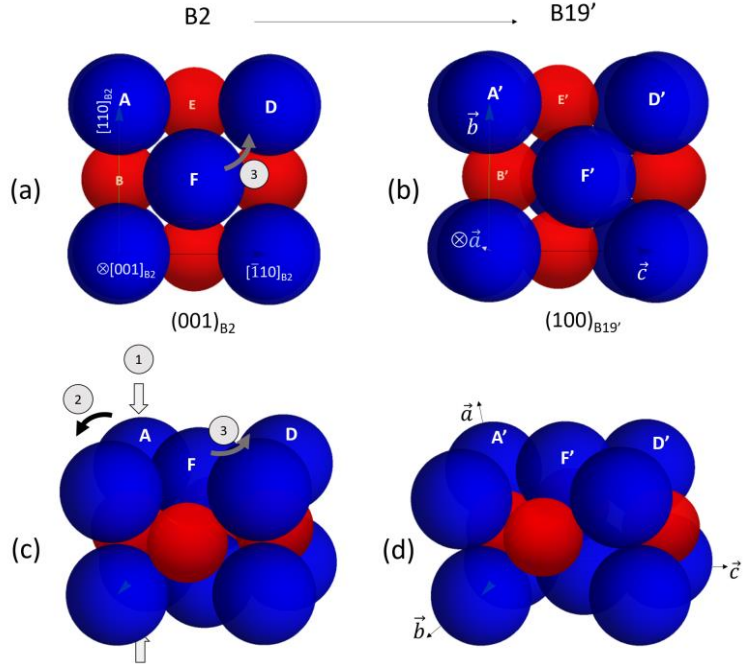

*Fig. 2. 3D representation of the B2 → B19' transformation, (a,b) viewed from the top along the direction [001]$_{B2}$, and (c,d) viewed sideways. The initial cubic B2 structure is in (a,c) and the final monoclinic B19' in (b,d). The three movements that occur simultaneously are given in the circles. In (b) the vector $a_{B19'}$ points toward the rear of the page but is not exactly perpendicular to the page because of the monoclinicity.*

The Ti atom initially in G in B2 is in G' = $\overrightarrow{OG'}$ = (x, y, z) in B19'. Its coordinates are determined numerically in the orthonormal basis $\mathcal{B}_{\boxplus}$ with Mathematica by solving the equations

$$\begin{cases} \|\overrightarrow{E'G'}\| = \|\overrightarrow{OG'} - \overrightarrow{OE'}\| = d_{TiNi} \\ \|\overrightarrow{C'G'}\| = \|\overrightarrow{OG'} - \overrightarrow{OC'}\| = d_{Ti} \\ \|\overrightarrow{E'F'}\| = \|\overrightarrow{E'G'} + \overrightarrow{OA'}\| = d_{TiNi} \end{cases} \quad (6)$$

In $\mathcal{B}_{\boxplus}$, the coordinates of the Ni atom in E' are given by equation (3), those of the Ti atom in C' are simply $\overrightarrow{OC'} = (0,0,c_{B19'})$, and those of the Ti atoms in F' are deduced from $\overrightarrow{E'F'} = \overrightarrow{E'G'} + \overrightarrow{OA'}$, with $\overrightarrow{OA'} = [\mathcal{B}_{\boxplus} \to \mathcal{B}_{B19'}]\begin{bmatrix}1\\0\\0\end{bmatrix} = (d_{Ti}\sin\beta_{B19'}, 0, d_{Ti}\cos\beta_{B19'})$. Once the coordinates of the Ti atoms in F' and G' are determined, those of the Ni atom in B' are deduced by $\overrightarrow{G'B'} = \overrightarrow{E'A'}$, or equivalently $\overrightarrow{OB'} = \overrightarrow{OG'} + \overrightarrow{OA'} - \overrightarrow{OE'}$.

The equations (6) are solved numerically. The y-coordinates of the Ti atoms in F' and G' and those of the Ni atoms in E' have the same value $y_{G'}$. Since the vector $\boldsymbol{b}_{B19'}$ is normal to the plane $(010)_{B19'} \parallel (110)_{B2}$, this y-value is half of the norm of $\boldsymbol{b}_{B19'}$. It comes that:

$$b_{B19'} = 2\,y_{G'} = 4.147 \text{ Å} \quad (7)$$

## 2.4 Orientation and distortion matrices

Now that the parameters $\left(a_{B19'}, b_{B19'}, c_{B19'}, \beta_{B19'}\right)$ are known, $[\mathcal{B}_\boxplus \to \mathcal{B}_{B19'}]$, the structure tensor of B19', can be calculated. In addition,

$$[\mathcal{B}_\boxplus \to \mathcal{B}_{B2}] = a_{B2} \begin{pmatrix} 0 & 0. & 1 \\ 1/\sqrt{2} & 1/\sqrt{2} & 0 \\ -1/\sqrt{2} & 1/\sqrt{2} & 0 \end{pmatrix} \tag{8}$$

The orientation relationship matrix between B2 and B19' then comes directly as

$$\boldsymbol{T}^{B2 \to B19'} = [\mathcal{B}_\boxplus \to \mathcal{B}_{B2}]^{-1}[\mathcal{B}_\boxplus \to \mathcal{B}_{B19'}] = \begin{pmatrix} 0.0936 & 0.9742 & -1.0896 \\ -0.0936 & 0.97421 & 1.0896 \\ 0.9543 & 0. & 0. \end{pmatrix} \tag{9}$$

The distortion matrix from B2 to B19' is

$$\boldsymbol{F}^{B2} = \boldsymbol{T}^{B2 \to B19'} \boldsymbol{C}^{B19' \to B2} = \begin{pmatrix} 1.0319 & -0.0577 & 0.0936 \\ -0.0577 & 1.0319 & -0.0936 \\ 0. & 0. & 0.9543 \end{pmatrix} \tag{10}$$

Its determinant gives the molar volume change of the transformation,

$$\frac{\mathcal{V}_{B19'}}{\mathcal{V}_{B2}} = Det(\boldsymbol{F}^{B2}) = 1.0130 \tag{11}$$

This means that a volume change of +1.3% is expected.

## 2.5 Atomic positions in the hard-sphere B19' structure

The coordinates of the Ti atoms in F' and G' that were calculated in the basis $\mathcal{B}_\boxplus$ can be written in the conventional crystallographic basis $\mathcal{B}_{B19'}$ thanks to the structure tensor (4), similarly as what was made for the coordinates of the Ni atom in E' in equation (5). The atomic positions in the B19' structure deduced from the hard-sphere model are reported in the last column of Table 1. The atomic positions noted "Ti1" and "Ti2", and "Ni1" and "Ni2" are equivalent because of the $2_1$ screw symmetry axis. It is thus possible to write the structure more simply by shifting the lattice origin onto the $2_1$ axis as recommended by International Table for Crystallography (ITC) for the monoclinic $P2_1/m$ structure. In this way, only the coordinates of one of the two Ti atoms and one of the two Ni atoms are required, those of the second atoms are deduced by the Wyckoff equivalencies. The new coordinates of the Ti and Ni atoms in the ITC basis are given in the last row of Table 1.

Table 1. Lattice parameters and atomic positions of NiTi B19' structure. First row, coordinates given in the conventional crystallographic basis of B19' with a Ti atom at the origin O. Ti1 designates the Ti atom in position O, Ti2 in G', Ni1 the Ni atom in E' and Ni2 in B' in Fig. 1f. Second row, coordinates given in the ITC crystallographic basis of B19', with the $2_1$ screw axis passing through the new origin. The two first columns report some structures deduced from X-ray diffraction, the third one the structure predicted by DFT, and the last one the structure resulting from the hard-sphere model with inputs $d_{Ti} = 2.9$ Å, $d_{Ni} = 2.27$ Å, and $\beta = 97.9°$.

|  |  | Michal & Sinclair 1981, X-ray [3] | Kudoh et al. 1985, X-ray [5] | Huang et al. 2003, DFT [11] | Hard-sphere model (present work) |
|---|---|---|---|---|---|
| **Lattice parameters** | | $a$ = 2.885 Å<br>$b$ = 4.120 Å<br>$c$ = 4.622 Å<br>$\beta_{B19'}$ = 96.8° | $a$ = 2.898 Å<br>$b$ = 4.108 Å<br>$c$ = 4.646 Å<br>$\beta_{B19'}$ = 97.78° | $a$ = 2.929 Å<br>$b$ = 4.048 Å<br>$c$ = 4.686 Å<br>$\beta_{B19'}$ = 97.8° | $a$ = 2.9 Å<br>$b$ = 4.147 Å<br>$c$ = 4.638 Å<br>$\beta_{B19'}$ = 97.9° |
| **Atomic positions conventional basis of B19'** | T1 | (0, 0, 0) | (0, 0, 0) | not given | (0, 0, 0) |
| | Ti 2 | (0.055, ½, 0.558) | (0.1648, ½, 0.5672) | | (0.1277, ½, 0.5811) |
| | Ni1 | (0.580, 0, 0.472) | (0.6196, 0, 0.4588) | | (0.6024, 0, 0.4658) |
| | Ni2 | (0.475, ½, 0.086) | (0.5452, ½, 0.1084) | | (0.5253, ½, 0.1152) |
| **Atomic positions in the ITC B19' basis** | Ti | (0.4726, ¼, 0.221) | (0.4176, ¼, 0.2164) | (0.4122, ¼, 0.2173) | (0.4361, ¼, 0.2095) |
| | Ni | (0.0525, ¼, 0.693) | (0.0372, ¼, 0.6752) | (0.0469, ¼, 0.6755) | (0.0385, ¼, 0.6753) |

# 3 Discussion

## 3.1 Comparison of the hard-sphere B19' structure with the literature

The output lattice parameters $a_{B19'}, b_{B19'}, c_{B19'}$ deduced from the hard-sphere are close to those reported by Kudoh *et al.* [5] from X-ray diffraction; the differences are only $+0.07\%, +0.9\%, -0.2\%$, respectively. The differences of the atomic positions can be evaluated by calculating in the orthonormal basis the distance between the hard-sphere positions and those reported in Ref. [5]; they are 0%, 3.9%, 1.8%, 2.1% for Ti1, Ti2, Ni1, Ni2, respectively. The deviation with other studies is larger. In Fig. 3, the X-ray theoretical diffractograms simulated from different B19' structures reported in various studied are plotted in green; the experimental X-ray diffractogram of B19' obtained by Wang *et al.* [2] is in black. The simulations were performed with the open source software *Dans_Diffraction* developed by Dan Porter [24] from the cif files of the phases [25]. None of the simulations fits perfectly with the experiment, but some structures look worse than others; for example, the B19' structure resulting from DFT calculations by forcing the monoclinicity at 97.8° (instead of 107° that gives the minimum of energy) [11] generates peaks on the diffractogram that do not coincide with any of the experimental ones (Fig. 3c). The agreement is better with the structure proposed by Kudoh *et al.* [5] (Fig. 3b) or with that deduced from the hard sphere model (Fig. 3d). Note that some peaks in the experimental diffractogram are broad, and some of them such as the (002) and (111) appear nearly split. We think that this broadening is not related to a triclinic tendency of the B19' structure, but, as mentioned in introduction, is the consequence of the elastic field that accommodate the incompatibilities between the twin-related variants [8].

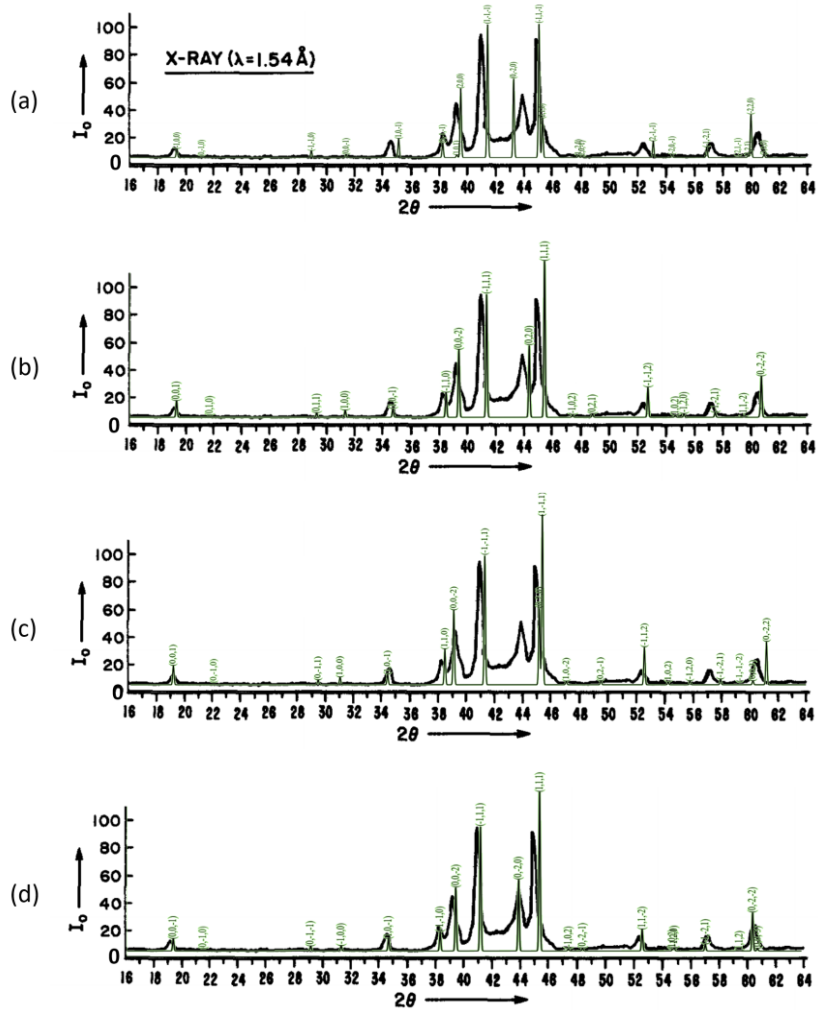

*Fig. 3.   Experimental X ray diffractogram of B19' structure from Ref.[1] in black, with theoretical lines in green simulated with Dans_Diffraction software from the structures proposed by (a) Michal and Sinclair 1981 [3], (b) Kudoh et al. 1985 [5], (c) Huang et al. 2003 [11], and (d) from the present hard-sphere model.*

## 3.2   Possible continuous paths from B2 to B19'

The extreme simplicity of the hard-sphere model permits to capture easily the three main components of the lattice distortion, as illustrated in Fig. 2c,d. The movement 1 is the contraction along the direction $[001]_{B2}$ driven by the formation of a new Ti-Ti bond. The movement 2 is an angular distortion given by the opening of the angle between the $[\bar{1}10]_{B2}$ and $[110]_{B2}$ direction from 90° to the monoclinic angle $\beta_{B19'}$. During this movement, the Ni atom keeps contact with three of the four Ti atoms in the first layer of the $(110)_{B2}$ plane. The movement 3 is a rotation of the Ti atoms on the second layer of the $(110)_{B2}$ plane, as if they were rolling on the Ni atom of the first layer of $(110)_{B2}$. There is no unambiguous way to combine the three movements; different paths can be imagined. A very improbable one would be a sequential sequence of movements 1, 2, and 3. More probably the three movements occur simultaneously in time such that the lattice distortion from B2 to B19' is complete is a few ns. We call the lattice parameters of the intermediate states $(a, b, c, \beta)$. They are all function of time. The parameter $a$ is given by the movement 1; it varies from the initial state $a_i = a_{B2}$ to $a_f = a_{B19'}$. The parameter $\beta$ is given by the movement 2; it varies from $\beta_i = 90°$ to $\beta_f = \beta_{B19'} = 97.9°$. The parameter $c$ and the position of the Ni atom in the first layer of $(110)_{B2}$ are outputs of calculations. The position of the Ti atom in the second layer varies continuously by the movement 3 from F to F'. The parameter $b$ is an output of the calculations. Instead of time, it is possible to use the angle $\beta$ as the varying parameter. The movement 2 becomes a simple linear function of $\beta$. One can then propose that the parameter $a$ is a power function of $\beta$, i.e. $a(\beta) = a_i + (a_f - a_i)(\frac{\beta-\beta_i}{\beta_f-\beta_i})^n$, where the exponent $n$ is a positive number arbitrarily chosen. If $n = 1$, $a$ is a linear function of $\beta$. Note that for an exponent value $n \approx 0$, the evolution of $a(\beta)$ is fast at the early stages of the transformation, when $\beta \approx \beta_i$. At the opposite, if the exponent is $n \gg 1$, $a(\beta)$ nearly does not change for values $\beta \approx \beta_i$. Once the law $a(\beta)$ is chosen, the hard-sphere model permits to calculate the positions of the atoms in the first layer and the value of $b(\beta)$. It also gives the positions of the atoms in the second layer and the value of $b_{max}(\beta)$ if the movement 3 is complete, i.e. when the Ti atoms of the second layer initially in F comes into contact with the atom D of the first layer. Since there is no reason to believe that these atoms come immediately into contact, we can imagine that the intermediate states of the movement 3 are given by a power function of $\beta$ of exponent $m$, $b(\beta) = b_i + (b_{max}(\beta) - b_i)(\frac{\beta-\beta_i}{\beta_f-\beta_i})^m$. A movie showing the atomic displacements during the B2 → B19' transformation can be made for different paths, as illustrated in Fig. 4 in the case $n = 1$, $m = 1$ (linearity).

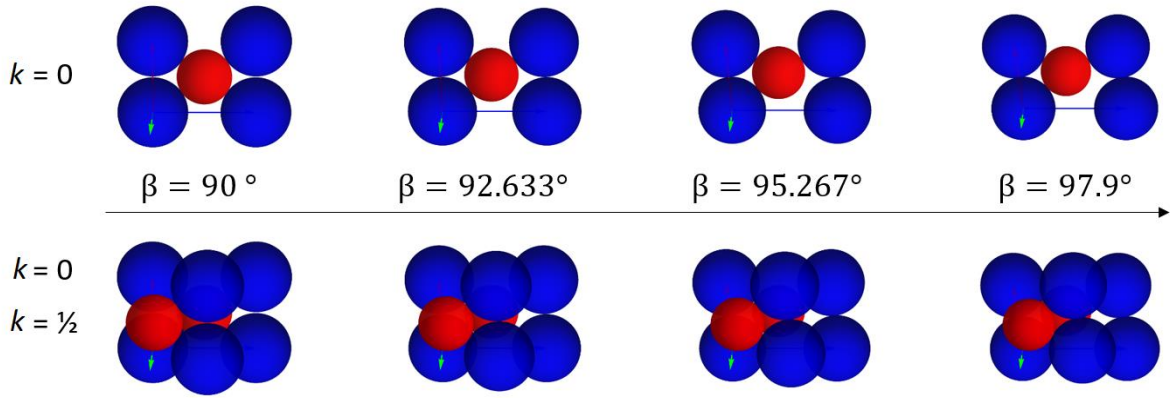

*Fig. 4. Continuous transformation path from B2 (β = 90°) to B19' (β = 97.9°) in the case n = 1 and m= 1, illustrated by four snapshots. The top row shows the 1st layer k = 0 of the (110)B2//(010)B19' plane, and the bottom row the 1st layer k= 0 overlapped by the 2d layer k = ½.*

The volume change $\frac{V'}{V_{B2}}$ during the transformation is given by the determinant of the distortion matrix calculated at each step $\beta$. The curves corresponding to different paths ($n$, $m$) are shown in Fig. 5. Note that the curves starts at a value lower than 1 because the diameter of the Ni atom in the hard-sphere model are smaller than in the actual B2 structure. Indeed, the Ti-Ni distance is $d_{TiNi} = \frac{\sqrt{3}a_{B2}}{2} = 2.607$ Å for B2 whereas it is $d_{TiNi} = \frac{d_{Ni}+d_{Ti}}{2} = 2.585$ Å. The volume drop due to the brutal change of atomic size imposed by the model is $\left(\frac{2\,d_{NiTi}/\sqrt{3}}{a_{B2}}\right)^3 \approx 0.975$. From this starting value, most of the curves increases monotonically with $\beta$, as shown by the dotted curves. Local minima or maxima exist only for exponents $(n, m)$ close to zero or largely higher than 1. In the case $(n = 0.1, m = 1)$ shown by the blue solid curve in Fig. 5a, the volume ratio decreases below 0.975 at the early stage of the transformation because of the sudden contact established between the Ti atoms in O and A. This minimum volume ratio is $\frac{a_{B19'}}{a_{B2}} \approx 0.96$. In the cases $(n = 10, m = 1)$ or $(n = 1, m = 10)$ shown by the red solid curves in Fig. 5a and Fig. 5b, respectively, a local maxima is reached for values in the range $\beta \approx 96° - 97°$ because the contact between the Ti atoms (in O and A for the case $n = 10$, or in F and D for the case $m = 10$) is established only at the end stage of the transformation.

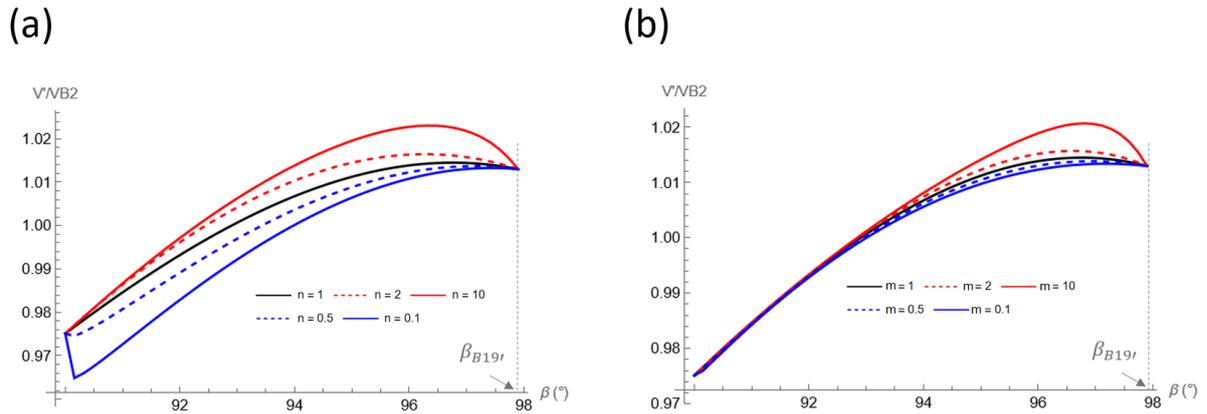

Fig. 5. Volume change $\frac{V'}{V_{B2}}$ calculated by the hard-sphere model during the transformation, between the initial state that is like B2 but with a smaller atomic size for Ni atoms ($\beta$ = 90°) and the final state B19' ($\beta_{B19'}$ = 97.9°). Different possible paths are explored: (a) for various exponent n in the power law of a($\beta$) and fixing m = 1 for b($\beta$), (b) for various exponent m in the power law of b($\beta$) and fixing n = 1 for a($\beta$).

Since the instant of the change of diameter of the Ni atoms is not necessarily at the early beginning of the transformation and is actually not known (if such a change have a physical meaning), and since this contribution is important in the total volume change, the curves shown in Fig. 5 should be taken with caution.

### 3.3 An explanation to the value of the monoclinic angle?

The structure of the B19' has been built by assuming that the monoclinic angle is $\beta_{B19'} = 97.9°$. However, the hard-sphere model can be used to explore the hypothetical B19' structures with other monoclinic angles. The equations introduced in section 2 with different final $\beta_{B19'}$ values have been solved with the same Mathematica program as that used in section 2. The $\beta_{B19'}$ values were chosen from 90° to 112° with a step of 2°. The volume change $\frac{V_{B19'}}{V_{B2}}$ calculated for each final $\beta_{B19'}$ value follows a concave-down curve shown in Fig. 6. In order to get a better accuracy on the local maximum, a step of 1° has been used to refine the curve between 96° and 102°, and 0.1° between 98° and 99°. The maximum volume change is obtained for $\beta_{B19'} = 98.5°$, and the associated volume change of the B2 → B19' transformation is 1.0132.

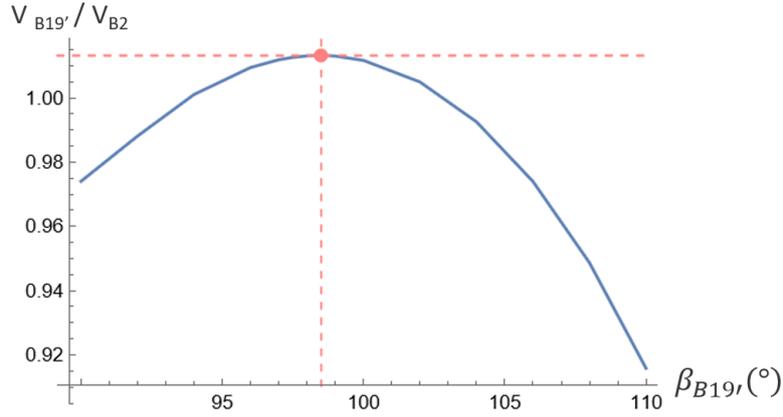

*Fig. 6. Volume change $\frac{V_{B19'}}{V_{B2}}$ calculated by the hard-sphere model for different hypothetical B19' structures depending on their monoclinic angle $\beta_{B19'}$*

It can be noted that the maximum is reached for a monoclinic angle very close to the actual value $\beta_{B19'} = 97.9°$. This fact may not be a coincidence, even if counterintuitive at first look. Indeed, if we consider that the B19' martensite should nucleate and grow inside the parent B2 phase, a small volume change could be expected for a better austenite/martensite accommodation. Thus, it is not a coincidence, the reason for the monoclinic value $\beta_{B19'} = 97.9°$ is probably not mechanical. An explanation can be proposed based of the entropy contribution to the Gibbs free energy. The molar entropy of a phase is constituted of electronic, configurational and vibrational contributions; and according to the literature, this last term is predominant in the total entropy of solids [26][27]. Since in rough approximation this term is a function of the mean squared displacements of the atoms, which themselves depends on the molar volume, a larger volume allows a larger entropy. An analogy can be taken with an ideal gas for which the molar entropy is a function of the volume by the equation $\Delta S_{V_1 \to V_2} = R \ln(V_2/V_1)$, where $\Delta S_{V_1 \to V_2}$ is the increase of entropy when a mole of the gas expands from the volume $V_1$ to the volume $V_2$. At the opposite, the molar enthalpy is probably less dependent of $\beta_{B19'}$ than the molar entropy. Indeed, according to the hard-sphere model, the number of short first neighbourhood Ti-Ti and Ti-Ni bonds between the atoms in B19' do not depend on $\beta_{B19'}$ as long as $\beta_{B19'} \neq 90°$; thus the molar enthalpy is a function of $\beta_{B19'}$ only by the contribution of the second neighbour bonds. The hypothesis that the value of $\beta_{B19'}$ is imposed by a maximum molar volume is quite speculative at the moment and would require a study of the phonon dispersions in the structure. However, if it is correct, it would imply that the value of $\beta_{B19'}$ would not result from the accommodation stress field between the parent B2 and B19' as proposed by Ishii [15]. It would also mean that $\beta_{B19'}$ could have been determined from the hard-sphere model itself without requiring it as an input. Unfortunately, we could not check the validity of the hypothesis on other B19' structures of other alloys, because to our knowledge B19' only exists in NiTi.

### 3.4 Extrapolation of the hard-sphere model to orthorhombic B19 structures

In section 3.2, the starting structure used in the continuous model of B2 to B19' transformation is B2 ($\beta = 90°$) except that the Ni atoms were chosen already slightly smaller than in the real B2 phase. None of the three movements of atoms were not yet activated at the beginning of the transformation process. Let us now consider a different problem in which $\beta = 90°$, but the movements 1 and 3 are complete. This structure would be a type of B19' structure without monoclinic distortion; such a structure is expected to be B19. Can the hard-sphere model used for B19' could also be applied to predict the lattice parameters and atomic positions of the B19 phase? Before comparing with the literature, some important crystallographic conventions have to be reminded. Contrarily to B19' which exists only in NiTi alloys, the B19 phase has been reported in various binary alloys, such as AgCd, AuCd, AuTi, CdMg, IrW, MoPt, NbPt, NbRh, PtTi, PtV [28]-[32]. Its space group is Pmma (n° 51). It is characterized by a unit cell made of two atoms A and B, with A in Wyckoff positions *e* (1/4, 0, *z*) and (3/4, 0, -*z*) and B in Wyckoff position *f* (1/4, 1/2, *z*) and (3/4, 1/2, -*z*). Note that the structure is invariant if the atom pairs A and B are exchanged, i.e. if the atoms A are placed in positions *f* and the atoms B in positions *e*. In order to directly compare with the literature the "hard-sphere" B19 structure imagined as a special case of B19' with $\beta$ = 90°, the $2_1$ axis positioned along the **b**-axis in B19' should become the glide direction **a** of B19, and the **c**-axis should be reversed in order to keep a positive handedness. In other words, the vectors (**a**, **b**, **c**) of B19' described in the previous sections should become the vectors (**b**, **a**, -**c**) of B19. For the atomic positions, we also have to apply the operation (*x*, *y*, *z*)$_{B19'}$ ⇒ (*y*, *x*, -*z*)$_{B19}$.

If we assume that the B19 structure is a special type of B19' without monoclinicity, and keep the same size of the Ti and Ni atoms as those used for the monoclinic B19' structure, i.e. $d_{Ti} = 2.90$ Å and $d_{Ni} = 2.27$ Å, the lattice parameters of B19 would be *a* = 4.26 Å, *b* = 2.9 Å, *c* = 4.28 Å. The *c* value does not agree well with that measured by X-ray diffraction in the B19 phase reported in NiTiCu alloys (*a* = 4.28 Å, *b* = 2.88 Å, *c* = 4.51 Å) [29]. If we make the hypothesis that the Ni atoms have their metallic diameter $d_{Ni} = 2.45$ Å, the hard-sphere model would give *a* = 4.43 Å, *b* = 2.9 Å, *c* = 4.50 Å, which now would be perfect for the *c* value but not anymore for *a*. We have to conclude that the hard-sphere model is not satisfying to describe the B19 phase in NiTi alloys. However, the B19 structure is observed inthese alloys only when they contain at least 10% of Cu [29]; the complex electronic structure of copper may thus be at the origin of a strong directional bonding with the Ti atoms and make the hard-sphere model irrelevant. If this explanation is correct, one can wonder if the hard-sphere model of B19 can be applied to "simpler" binary structures. In order to explore this hypothesis, we considered three alloys in which the lattice parameters of B19 were determined by X-ray diffraction: AuTi, PdTi and AuCd. The atomic positions were also reported for the two last alloys [30][31]. We used the same equations as those described in section 2; we just slightly adjusted the size of the A and B atoms around their metallic diameter in order to obtain the best fits with the lattice parameters and the atomic positions reported in the literature. The results are presented in Table 2. They are in excellent agreement with the observations. Note that the PdTi and PtTi structures have nearly the same lattice parameters and atomic positions, that is why they are reported in the same column in Table 2. It can thus be concluded that in most of the binary alloys, the B19 structures are simple hard-sphere structures. The only exception seems to be in NiTi(Cu) alloys, probably because of the presence of copper.

Table 2. B19 structures calculated from the hard-sphere model, and compared with the literature. The diameters of the atoms (inputs), slightly adjusted from the metallic diameter of the elements, are indicated in the first row. The lattice parameters and atomic positions (outputs) are in the second and third rows, respectively. Note that for PdTi, the z positions in the Wyckoff positions e and f reported in Ref. [32] have been changed by 1-z in order to facilitate the comparison with the other studies.

|  |  | NiTi |  | AuTi |  | PdTi or PtTi |  | AuCd |  |
|---|---|---|---|---|---|---|---|---|---|
| **Atom diameter** | | hard-sphere<br>e : $d_{Ti}$ = 2.9 Å<br>f : $d_{Ni}$ = 2.45 Å | X ray [29] | hard-sphere<br>e : $d_{Ti}$ = 2.9 Å<br>f : $d_{Au}$ = 2.75 Å | X-ray [30] | hard-sphere<br>e : $d_{Ti}$ = 2.8 Å<br>f : $d_{Pd}$ = 2.77 Å | X-ray [30][32] | hard-sphere<br>e : $d_{Cd}$ = 3.15 Å<br>f : $d_{Au}$ = 2.62 Å | X-ray [31] |
| **Lattice parameters** | | $a$ = 4.43 Å<br>$b$ = 2.9 Å<br>$c$ = 4.50 Å | $a$ = 4.28 Å<br>$b$ = 2.88 Å<br>$c$ = 4.51 Å | $a$ = 4.65 Å<br>$b$ = 2.9 Å<br>$c$ = 4.85 Å | $a$ = 4.60 Å<br>$b$ = 2.93 Å<br>$c$ = 4.85 Å | $a$ = 4.56 Å<br>$b$ = 2.8 Å<br>$c$ = 4.82 Å | $a$ = 4.55 Å<br>$b$ = 2.78 Å<br>$c$ = 4.86 Å | $a$ = 4.78 Å<br>$b$ = 3.15 Å<br>$c$ = 4.83 Å | $a$ = 4.766 Å<br>$b$ = 3.151 Å<br>$c$ = 4.859 Å |
| **Atomic positions in B19 basis (Pmma)** | e | (¼, 0, 0.793) | Not reported | (¼, 0, 0.821) | Not reported | (¼, 0, 0.831) | (¼, 0, 0.82) | (¼, 0, 0.788) | (¼, 0, 0.7997) |
|  | f | (¼, 0, 0.293) |  | (¼, 0, 0.321) |  | (¼, ½, 0.331) | (¼, ½, 0.32) | (¼, ½, 0.288) | (¼, ½, 0.295) |

# 4  Conclusion

A hard-sphere model of the B2 → B19' transformation and of the B19' structure in NiTi alloys is proposed. The model's inputs are the Ti and Ni diameters and the monoclinic angle $\beta_{B19'}$. The transformation is decomposed into three distinct but concomitant atomic movements:

1. A contraction along the $[100]_{B2}$ axis resulting from the formation of short Ti-Ti bonds.
2. An angular distortion between the $[001]_{B2}$ and $[1\bar{1}0]_{B2}$ directions increasing their angle from 90° to $\beta_{B19'}$.
3. An rotational displacement of the Ti atoms around the Ni atoms resulting from the formation of short Ti-Ti bonds.

The outputs of the model are the lattice parameters and the final atomic positions in the B19' phase. They agree well with the structural data deduced from X-ray diffraction or DFT simulations reported in the literature. The differences on the lattice parameters are less than 1% and the differences on the atomic positions better than 4%.

The atomic trajectories during the B2 → B19' are not anymore viewed as "straight" shears and shuffles, but more as angular/rotational collective displacements driven by the formation of short Ti-Ti bonds. An important result is that the actual monoclinic angle $\beta_{B19'} \approx 97.9°$ has the highest molar volume among all the possible hypothetical hard-sphere monoclinic structures. If it is not a coincidence, the movement 2 (monoclinicity) could result from a maximization of the vibrational entropy. More research is required to explore this hypothesis.

The hard-sphere model was applied to the B19 structure by fixing $\beta \approx 90°$, and applying only the atomic movements 1 and 3. The calculations do not agree well with the literature for NiTi(Cu) alloys, probably because of the high amount of copper in these alloys. The model's outputs are however in excellent agreement with the B19 structures reported in other binary alloys such as AuTi, PdTi, and AuCd.

From this study, it is concluded that the B19' phase in NiTi alloys and the B19 phase in AuTi, PdTi, and AuCd alloys can be considered as simple hard-sphere structures.

**Acknowledgments:** The author would like to thank the Swiss National Science Foundation for indirectly supporting this study by financing the project on shape memory alloys (SNSF, No 200021_200411).